\documentclass[floatfix,showpacs,amsmath,amsfonts,amssymb,aps,twocolumn,prl,10pt]{revtex4-1}
\usepackage[pdftex]{graphicx}

\newcommand{\figref}[1]{Fig.~\ref{#1}}
\newcommand{\e}[1]{\text{e}^{#1}}

\newcommand{\diffd}{\text{d}}
\renewcommand{\vec}[1]{\mathbf{#1}}
\newcommand{\punc}[1]{\,#1}

\newcommand{\bra}[1]{\langle #1 |}
\newcommand{\ket}[1]{| #1 \rangle}
\newcommand{\overlap}[2]{\langle #1 | #2 \rangle}
\newcommand{\pos}{\vec{r}}
\newcommand{\Pos}{\vec{R}}
\newcommand{\ann}{c}
\newcommand{\cre}{c^\dagger}

\newcommand{\kf}[0]{k_\mathrm{F}}
\newcommand{\Ef}[0]{E_\mathrm{F}}
\newcommand{\Emax}[0]{E_\mathrm{max}}
\newcommand{\re}[0]{\mathrm{e}}

\begin{document}

\title{High-fidelity contact pseudopotentials and p-wave superconductivity}
\author{P.O.~Bugnion}
\author{R.J.~Needs}
\author{G.J.~Conduit}
\affiliation{Cavendish~Laboratory, J.J.~Thomson~Avenue, Cambridge, CB3~0HE, United~Kingdom}
\date{\today}

\begin{abstract} 
  We develop ultratransferable pseudopotentials for the
  contact interaction that are 100 times more accurate than contemporary
  approximations.  The pseudopotential offers scattering properties very
  similar to the contact potential, has a smooth profile to accelerate
  numerics by a factor of up to 4,000, and, for positive scattering lengths, does
  not support an unwanted bound state. We demonstrate these advantages in a
  Diffusion Monte Carlo study of fermions with repulsive interactions,
  delivering the first numerical evidence for the formation of a p-wave
  superconducting state.
\end{abstract}


\maketitle

Interparticle interactions 
are central to our understanding of correlated phenomena, but the ubiquitous
Coulomb and contact interparticle potentials diverge on coalescence,
impeding leading numerical methods. The contact interparticle potential is
realized in both ultracold atomic gases and idealized screened electrons,
making it an ideal testbed for developing ultratransferable
pseudopotentials. Common approximations to the contact interaction
display incorrect variations in the scattering phase shift with incident
particle energy and can harbor undesired bound
states~\cite{Astrakharchik04i,Astrakharchik04ii,Conduit09,Pilati10,
  Chang10,Giorgini99,Bugnion13,Bugnion13i,Bugnion13ii,Pilati10,Chang10}. We
develop a formalism for generating a bespoke pseudopotential for the contact
interaction that offers accurate scattering properties and has no
superfluous bound states. These advantages allow us
to deliver the first numerical evidence for a p-wave superconducting
instability in a fermionic gas with repulsive interactions.

The contact interaction is characterized by a scattering length $a$ that
parameterizes the variation of the scattering phase shift with incident
energy. The contact interaction comes in three flavors: sufficiently deep to
trap a two-body bound state ($a>0$), weakly attractive with no bound state
($a<0$), and repulsive ($a>0$). Contemporary numerical simulations of both
the bound state and weak attractive interactions adopt
a finite ranged square well or P\"oschl-Teller interaction. These
simulations have delivered crucial insights into the BEC-BCS
crossover~\cite{Astrakharchik04i,Morris10},
Bose gases~\cite{Astrakharchik04ii}, and few atom
physics~\cite{Bugnion13,Bugnion13i,Bugnion13ii}. However, the finite range
imbues the potential with incorrect scattering properties. While reducing
the range of the potential alleviates this, it slows numerical calculations.
The third category of contact potentials gives repulsive
interactions
that drive
itinerant ferromagnetism 
in Fermi gases~\cite{Conduit09,Pilati10,Chang10}, a Tonks-Girardeau
gas~\cite{Astrakharchik04ii}, and a Bose gas~\cite{Giorgini99}. 
The repulsive interaction is the first excited state of the bound state
potential so both have $a>0$, but in ultracold atomic gas
experiments~\cite{Jo09} the upper branch is protected by a slow three-body
loss process. To simulate repulsive interactions the first option is to
adopt a finite-ranged attractive potential~\cite{Pilati10,Chang10}. However,
to avoid forming the bound state, the trial wave function is restricted to
the Hartree-Fock excited state solution with no variational parameters,
leading to a poor estimate of the energy. Alternatively, one can adopt a
repulsive top-hat potential~\cite{Conduit09} potential with no bound state.
However, this 
has a finite range
greater than the scattering length,
resulting in an incorrect scattering phase shift.
The difficulty of simulating repulsive interactions means that there are
important open questions about fermionic gases: is the ground state of a
strongly interacting fermionic system
ferromagnetic~\cite{Conduit08,Conduit09,Conduit09i,Zhai09,Duine05,Maslov09,Chubukov09,Pedder13};
is the ferromagnetic transition first or second order; and whether
exotic phases that emerge around quantum criticality include a spin
spiral~\cite{Conduit09}, nematic phase~\cite{Chubukov09,Chubukov10,Karahasanovic12}, and a
counterintuitive p-wave superconductor~\cite{Balian:1963ve}. The p-wave
superconductor was suggested by perturbation
theory~\cite{Fay:1980kx,Mathur1998,Roussev:2001ys,Kirkpatrick:2001vn,Wang:2001kl,Conduit13}
and has been observed in
experiment~\cite{Saxena00,Huxley01,Watanabe02,Aoki01,Huy07,Gasparini:2010,Hattori12},
but has never been seen by numerics.

We develop a general ultratransferable
pseudopotential for the contact interaction. 
We test its accuracy using the exactly
soluble system of two trapped atoms, verify the first order itinerant
ferromagnet transition, and finally present the first numerical evidence for
a p-wave superconducting instability.

\section{Derivation of the pseudopotential}

\begin{figure}
 \includegraphics[width=1.0\linewidth,trim=0 0 0 0,clip=true]{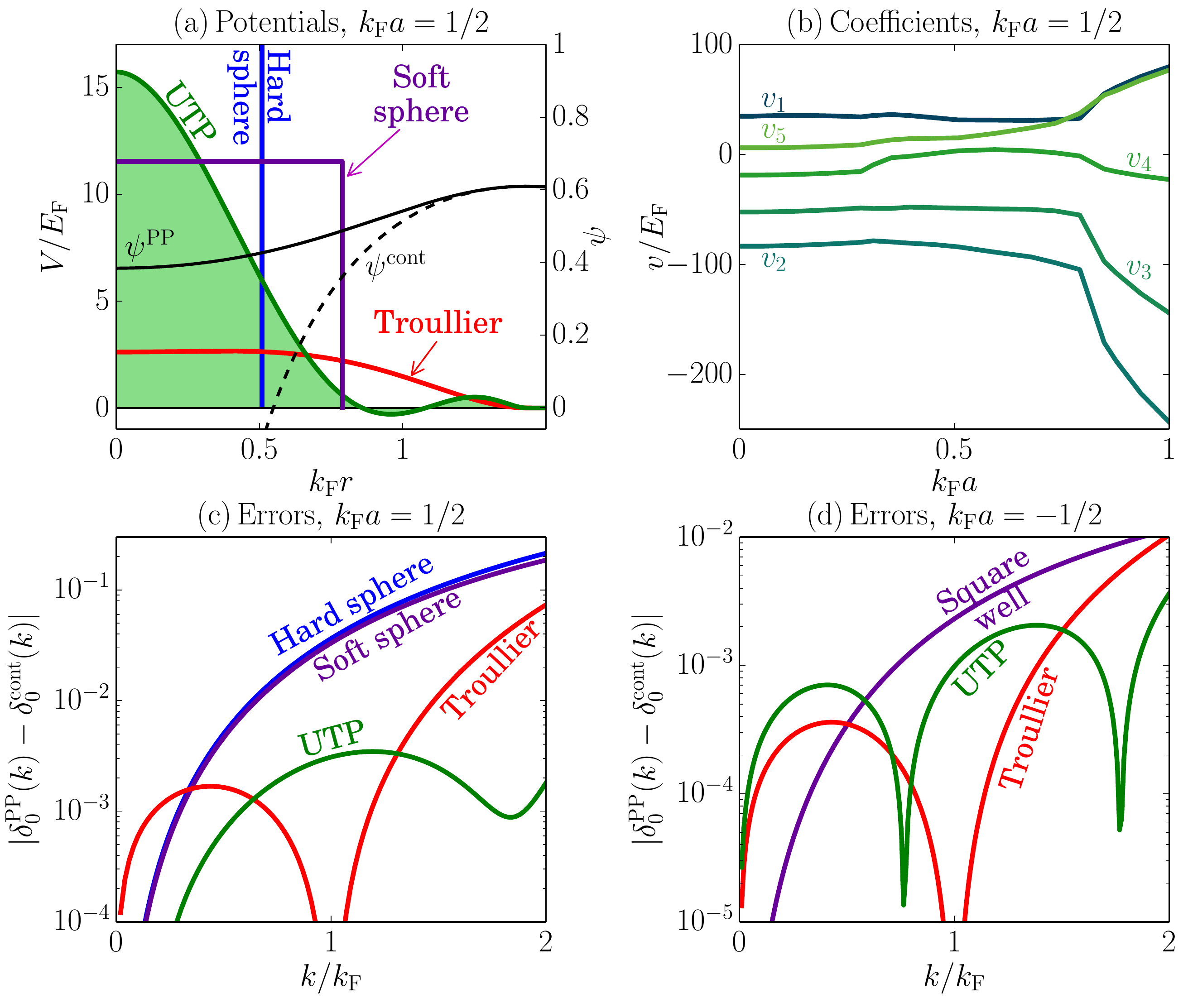}
 \caption{(Color online) (a) The pseudopotentials at $\kf a=1/2$
 on the repulsive branch. UTP denotes the ultratransferable pseudopotential.
 (b) The first five coefficients for the ultratransferable potential at $\kf a=1/2$.
 (c) The errors in phase shifts for the repulsive branch at $\kf a=1/2$. 
 (d) The errors in phase shifts for the attractive branch at $\kf a =-1/2$. 
 }
 \label{fig:phase_shifts}
\end{figure}

To construct the pseudopotential we study the two-body problem: two fermions
in their center-of-mass frame with wavevector $k\ge0$ and angular momentum
quantum number $\ell$. The Hamiltonian in atomic units ($\hbar=m=1$) is
$-\frac{\nabla^2}{2}\psi+V(r)\psi=\frac{k^2}{2}\psi$, with the contact
potential $V(r)=2\pi a\delta(r)\frac{\partial}{\partial r}
r$~\cite{Busch98}.  The scattering states are
$\psi^{\mathrm{cont}}_{k,\ell}=\sin[kr-\delta^\mathrm{cont}_\ell]/kr$, where $\delta_\ell$
is the scattering phase shift in angular momentum channel $\ell$. We seek a
pseudopotential that (i) reproduces the correct phase shifts over the range
of wavevectors $0<k\lesssim2\kf$, where $\kf$ is the Fermi wavevector, (ii)
supports no superfluous bound states to be compatible with ground state
methods and (iii) is smooth and broad to accelerate numerical calculations.
We first focus on positive scattering lengths $a > 0$, with no bound
state. We describe four families of pseudopotentials: hard sphere, soft
sphere (top hat), the Troullier-Martins form of norm-conserving
pseudopotentials~\cite{Troullier91,Hamann79} and the new ultratransferable pseudopotential.

The usual approach~\cite{Pilati10,Conduit09} starts from the low energy
expansion for the $s$-wave scattering phase shift
$\cot\delta_0=-\frac{1}{ka}+\frac{1}{2}kr_\mathrm{eff}+\mathcal{O}(k^3)$
where $r_\mathrm{eff}$ is the ``effective range'' of the potential.  For a
contact potential, $r_\mathrm{eff}$ and all higher order terms are zero.
Perhaps the simplest pseudopotential is a hard sphere potential with radius
$a$. This reproduces the correct scattering length $a$, thus delivering the
correct phase shift for $k=0$. However, the hard sphere has an
effective range $r_\mathrm{eff}=2a/3$. \figref{fig:phase_shifts}(c) shows
that this causes significant deviations in scattering power for $k>0$.

To improve the scattering phase shift, Ref.~\cite{Conduit09} adopted a soft
sphere potential: $V(r)=V_0\Theta(r-R)$, with $V_0$ and $R$ chosen to
reproduce the correct scattering length $a=R(1-\tanh\gamma/\gamma)$ and
effective range $r_\mathrm{eff}=R[1+\frac{3\tanh\gamma-\gamma(3+\gamma^2)}
{3\gamma(\gamma-\tanh\gamma)^2}]=0$, where $\gamma=R\sqrt{2V_0}$. The first
two terms in the low energy expansion of the phase shift are now correct,
leading to a small reduction in phase shift error in
\figref{fig:phase_shifts}(c).

The two potentials considered so far display incorrect behavior for larger
wavevectors due to the focus on reproducing the correct $k=0$ scattering
behavior. To improve the accuracy we turn to the
Troullier-Martins~\cite{Troullier91} formalism developed for constructing
attractive electron-ion pseudopotentials. These pseudopotentials reproduce
both the correct phase shift and its derivative with respect to energy at a
prescribed calibration energy (when constructing an electron-ion
pseudopotential, this is the bound state energy in an isolated
atom~\cite{Hamann79,Zunger79,Bachelet82,Hamann89,Rappe90,Lin93}). By
calibrating at the energy corresponding to the median incident scattering
wavevector $k=\kf$, we reduce the errors in the scattering phase shift over
a broad range of wavevectors. This delivers the pseudopotential shown in
\figref{fig:phase_shifts}(a) that is smooth, leading to improved numerical
stability and efficiency.  \figref{fig:phase_shifts}(c) demonstrates that
this potential is exact at the calibration wavevector $k=\kf$ and delivers a
marked decrease in phase shift error across all wavevectors.

The three potentials deliver a significant progression in accuracy. The hard
sphere potential reproduces the correct scattering behavior at $k=0$.  Both
the soft sphere and Troullier-Martins potential are transferable: the former
 producing correct scattering around $k=0$ and the latter around $k\sim\kf$
The significant improvement delivered by the Troullier-Martins potential
encourages us to develop the formalism to propose an \emph{ultratransferable}
pseudopotential that produces accurate phase shifts over all of the
wavevectors occupied in a Fermi gas.

To develop ultransferable pseudopotentials we continue to focus
on the contact potential, though the methodology can be readily generalized
to other interparticle interactions. We construct a pseudopotential that is
identical to the contact potential outside of a cutoff radius $r_c$, but
inside has a continuous first derivative at both $r=0$ and $r=r_c$,
\begin{align}
 \frac{V(r)}{\Ef}\!=\!
  \begin{cases}\! \left(1\!-\!\frac{r}{r_c}\right)^{2}\!
\left[v_{1}\left(\frac{1}{2}\!+\!\frac{r}{r_c}\right)\!+\!
\displaystyle\sum_{i=2}^{N_{\text{v}}}v_{i}\left(\frac{r}{r_c}\right)^{i}\right]&\!r\!\le\!r_c\\
0&\!r>r_c\punc{,}\nonumber
  \end{cases}  
\end{align}
with $N_{\text{v}}=9$. We choose the cutoff radius to correspond to the
first anti-node of the true wavefunction
By choosing a cutoff that is beyond the first node in the wavefunction, we guarantee that the
pseudopotential will not harbor a bound state, as demonstrated in
\figref{fig:phase_shifts}(a).  We calculate the scattering solution
$\psi^\mathrm{PP}_{k,\ell}(r)$ of the pseudopotential numerically to determine
the phase shift $\delta_\ell^\mathrm{PP}(k(k))$. The
difference in the scattering phase shift $\delta_\ell$ of the potentials is
characterized by the mean squared error in the phase shifts at the
cutoff radius,
\begin{align}
 \langle(\delta_\ell^\mathrm{PP}-\delta_\ell^\mathrm{cont})^2\rangle=
    \int_{0}^{2\kf}\!\!\left[ \delta_\ell^\mathrm{PP}(k)-
        \delta_\ell^\mathrm{cont}(k)
\right]^{2}\!\!\diffd k\punc{,}\nonumber
\end{align}
that is integrated over all wavevectors $0\le k\le 2\kf$ of interest. The
integrand can be convolved with a density of states to emphasize $k$ values
of interest.  We seek the variational parameters $\{v_{i}\}$ that minimize
the deviation
$\langle(\delta_\ell^\mathrm{PP}-\delta_\ell^\mathrm{cont})^2\rangle$ to
determine the pseudopotential that delivers the best approximation for the
contact potential. As demonstrated in \figref{fig:phase_shifts}(c), this
potential delivers an error in $\delta_0$ of less than $10^{-3}$ for all
wavevectors $0\le k\le2\kf$ found in a Fermi gas, corresponding to an
improvement of two orders of magnitude over previously used
pseudopotentials.

The pseudopotentials constructed will have finite scattering amplitude in
the p-wave and higher angular momentum channels. The contact potential, by
contrast, scatters only in the s-wave channel $\ket{s}$. This can be
solved by using a non-local
pseudopotential~\cite{Kleinman82,Fahy90} $\hat{V}=\ket{s}V(r)\bra{s}$, where
$\overlap{\pos}{s}=\mathrm{Y}_0^0(\pos)$, with the spherical harmonic
$\mathrm{Y}_0^0$ centered on either of the interacting particles. This
potential only acts on the s-wave component of the relative wavefunction.
Additional accuracy could also be gained by using different projectors for
different energy ranges~\cite{Vanderbilt90,Blochl94}.

{\it Attractive branch}: We can use a similar procedure to
derive pseudopotentials for the attractive branch $a<0$. For the attractive
case, the cutoff can be arbitrarily reduced to generate a potential that
tends to the contact limit, at the cost of computational efficiency.  For
example, in Monte Carlo simulations, the sampling efficiency is
approximately proportional to $r_c^3$.  In \figref{fig:phase_shifts} we
adopt a cutoff $r_c=1/2\kf$, and compare to the square well potential with
cutoff $r_c=0.01\sqrt[3]{3\pi^2}/\kf$ in Ref.~\cite{Astrakharchik04i}. Both
the Troullier-Martins pseudopotential and the ultratransferable pseudopotential have
an average error approximately 10 times smaller than the square well
potential, but their larger cutoff allows them to be sampled 4,000 times
more efficiently.

{\it Bound state}: To construct a pseudopotential for the bound state (corresponding to $a>0$),
we follow the Troullier-Martins prescription~\cite{Troullier91}. We
calibrate the pseudopotential at the binding energy $E=-1/2a^2$. The cutoff
is constructed in the same manner as for the attractive branch, delivering a
similar improvement in efficiency.

\section{Atoms in a trap}

\begin{figure}
 \includegraphics[width=1.0\linewidth,trim=5 0 0 0,clip=true]{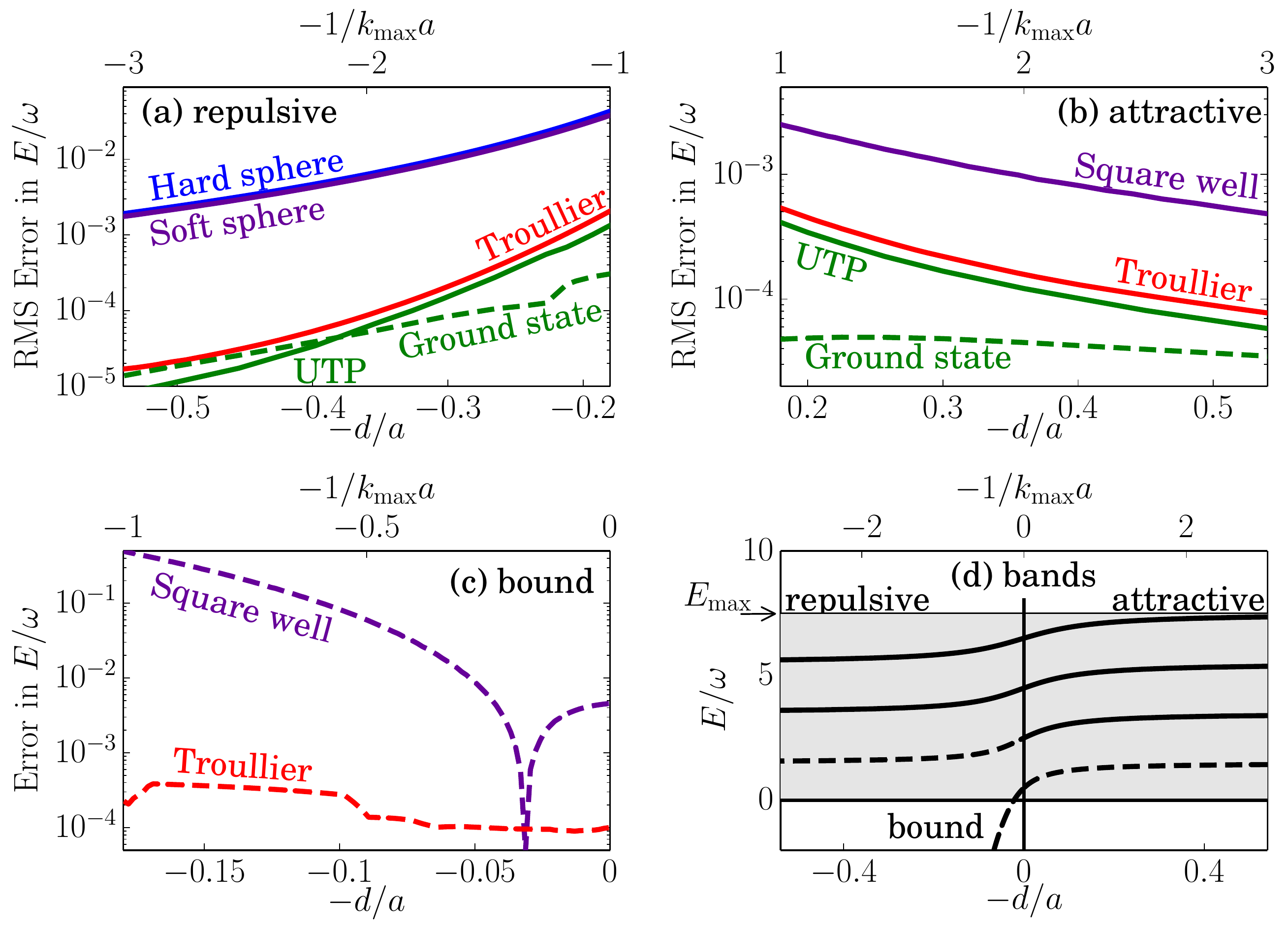}
 \caption{(Color online) Mean squared error in total energy for two atoms in
   a harmonic trap, for all bands below $\Emax$ (solid lines). (a) The error
   for repulsive interactions ($\kf a>0$). UTP denotes the ultratransferable
   pseudopotential. The dashed line denotes the error
   in the ground state energy with the ultratransferable pseudopotential. (b) The
   pseudopotential error for attractive interactions ($\kf a<0$). (c) The
   pseudopotential error in the bound state energy. (d) The band diagram for
   two atoms in a harmonic trap, calculated following Ref.~\cite{Busch98}.}
 \label{fig:busch_tests}
\end{figure}

We have developed a pseudopotential that delivers the correct scattering
phase shift for an isolated system. To test the pseudopotential we
turn to an experimentally realizable configuration~\cite{Serwane11,Zurn12}:
two atoms in a spherical harmonic trap with frequency $\omega$ and
characteristic length $d=1/\sqrt{\omega}$. For all three types of
interaction shown in \figref{fig:busch_tests}(d) this system has an
analytical solution~\cite{Busch98} that we can benchmark against, forming an
ideal test in an inhomogeneous environment. Moreover, the exact
solution extends to excited states, allowing us to test the performance of
the pseudopotential across a wide range of energy levels to provide a firm
foundation from which to study the many-body system.

\emph{Ground state}: We first compare the pseudopotential estimates of the
ground state energy to the exact analytical solution~\cite{Busch98}. For the
repulsive and attractive branches the hard/soft sphere potentials deliver
$\sim1\%$ error in the energy, whilst both the Troullier-Martins and
ultratransferable pseudopotentials (shown in \figref{fig:busch_tests}(a,b)) are
significantly more accurate with a $\sim0.01\%$ error. Finally we examine
the bound state energy in \figref{fig:busch_tests}(c). Both the square well
and Troullier-Martins formalism give the exact ground state energy for two
atoms in isolation. However, the trapping potential introduces
inhomogeneity, so the square well potential gives a $\sim10\%$ error in the ground
state energy, whereas the Troullier-Martins pseudopotential gives a
$\sim0.01\%$ error. This affirms the benefits of using a pseudopotential that is
robust against changes in the local environment. The success of the
Troullier-Martins and ultratransferable formalism at describing the ground
state is all the more significant considering these pseudopotentials aim to describe
the correct scattering properties over a range of energies. We would therefore
expect them to perform even better 
when modeling the excited states of the trap.

\emph{Excited states}: We now turn to examine the predictions for the
excited states in the repulsive and attractive branches. Due to the shell
structure, the excited states of a few-body system are related to the ground
state of a many-body system~\cite{Bugnion13}, allowing us to probe the
performance expected from the pseudopotential in a many-body setting. We
consider states up to a maximum energy $\Emax=7.5\hbar\omega$, corresponding
to 112 non-interacting atoms in the trap. In \figref{fig:busch_tests}(a,b)
the Troullier-Martins pseudopotential has a mean squared error averaged
over all bands below $\Emax$ that is between 10 and 100 times lower than
existing pseudopotentials. The ultratransferable pseudopotential is a further factor
of 2 more accurate. Additionally, when modeling the attractive branch, the
Troullier-Martins and the ultratransferable formalism are 4,000 times more efficient,
due to their larger cutoffs.  

\section{Repulsive fermions}

\begin{figure}
 \includegraphics[width=1.0\linewidth]{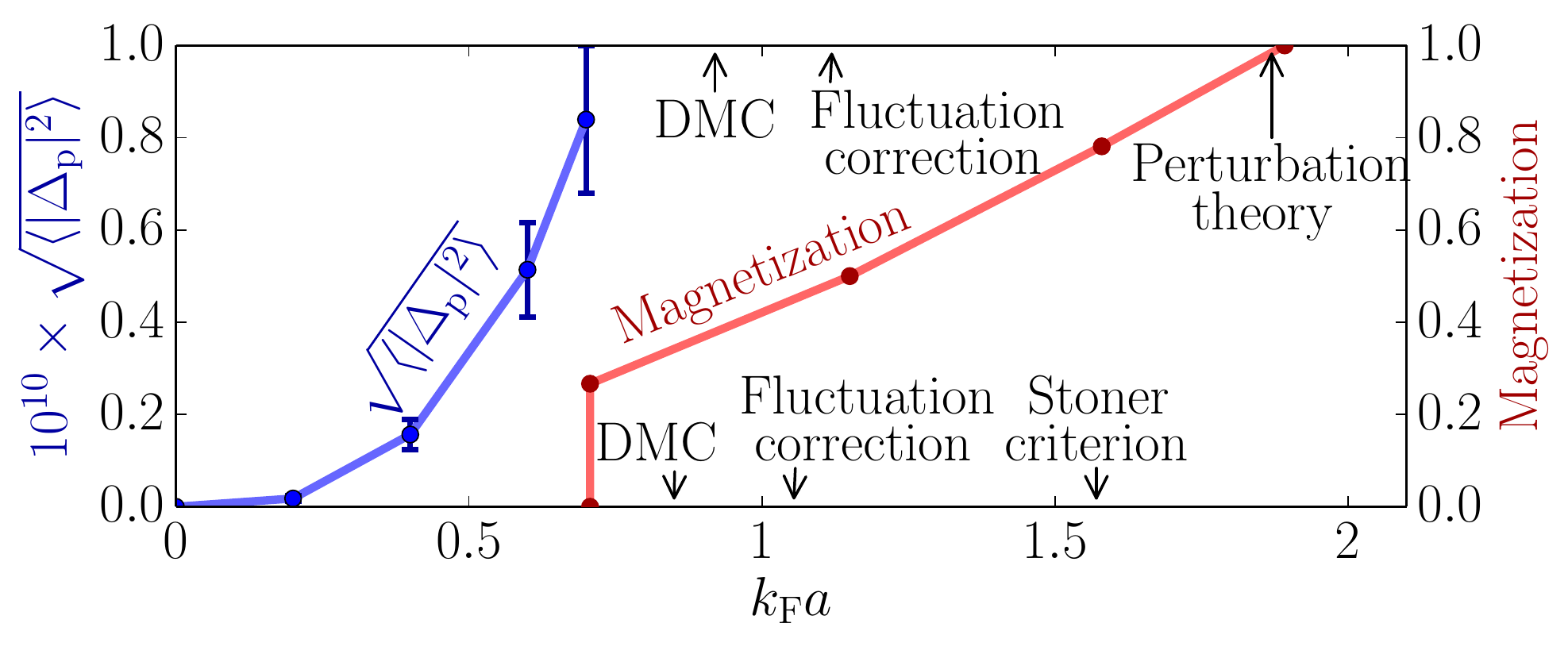}
 \caption{(Color online) The magnetic phase diagram and emergence of p-wave
   superconducting order. The error bars on the magnetization line are
   smaller than the markers. Labels show the predicted interaction strength
   for the onset of magnetization and entry into the fully polarized state
   from previous
   works~\cite{Duine05,Conduit08,Conduit09,Pilati10,Massignan11,Cui10}.}
 \label{fig:ferro}
\end{figure}

Having verified that the pseudopotentials reproduce the correct scattering
phase shift and bound state energy for two harmonically trapped atoms, we
now exploit their accuracy to study two unsolved questions in many-body ferromagnetic
metals tuned near quantum criticality: the nature of the ferromagnetic phase
transition and presence of p-wave superconducting correlations.

\emph{Quantum Monte Carlo}: We use fixed-node Diffusion Monte Carlo
(DMC)~\cite{Umrigar93} implemented in the
\textsc{casino} code~\cite{Needs10}, with a trial wavefunction $\Psi=\re^J
D_\uparrow D_\downarrow$, where $D_\alpha$ denotes a Slater determinant of
$N_\alpha$ plane waves. The Jastrow factor is taken to be
\begin{align}
J\!=\!\!\!\!\!\!\!\sum_{\substack{j\ne
    i\\\alpha,\beta\in\{\uparrow,\downarrow\}}}\!\!\!\!\!\!\!
 \left(1\!-\!\frac{|\pos_i\!-\!\pos_j|}{L^u_{\alpha\beta}}\right)^{\!\!2}\!\!
u_{\alpha\beta}(|\pos_i\!-\!\pos_j|)\Theta(L^u_{\alpha\beta}\!-\!|\pos_i\!-\!\pos_j|),\nonumber
\end{align}
where $u_{\alpha\beta}$ is a polynomial whose parameters we optimize in a
Variational Monte Carlo (VMC) calculation and $L^u_{\alpha\beta}$ is a cutoff
length~\cite{Drummond04}. We model spin polarized systems by performing
calculations for $N_\uparrow=81$ and $N_\downarrow\in\{81,57,33,27,19,7,1\}$
that correspond to filled shells. This guarantees that the trial
wavefunction is an eigenstate of the total spin operator $\hat{S}^2$ and the
spatial symmetry operators of the cubic lattice.

We use a backflow transformation~\cite{Chang10,Rios06} in the construction
of the orbitals that enter the Slater determinant, with the replacement
$\pos_{i\sigma} \to \pos_{i\sigma}+\sum{\begin{subarray}{l}j\ne
    i\\\alpha,\beta\in\{\uparrow,\downarrow\}\end{subarray}}
(\pos_i-\pos_j)\eta^{\alpha\beta}_{ij}(|\pos_i-\pos_j|)$ where
$\eta^{\alpha\beta}_{ij}(r)=(1-r/L^\eta_{\alpha\beta})^2
\Theta(L^\eta_{\alpha\beta}-r) p_{\alpha\beta}(r)$, $p_{\alpha\beta}$ is a
polynomial whose parameters are optimized in VMC, and $L^\eta_{\alpha\beta}$ is a
cutoff length. We reduce finite size effects by twist
averaging~\cite{Rajagopal94,Rajagopal95,Lin01} and correct the
non-interacting kinetic energy of the finite sized system with that of the
corresponding infinite system~\cite{Pilati10}.

\emph{Ferromagnetic phase transition}: In~\figref{fig:ferro} we observe a first order phase
transition to a partially polarized state at $\kf a=0.71$, markedly lower
than previous DMC predictions of $\kf a\sim0.85$ ~\cite{Pilati10,Conduit09}.
The system becomes fully polarized at $\kf a=1.89$, close to the theoretical
prediction of 1.87~\cite{Massignan11,Cui10}. This is significantly larger
than the values calculated previously using DMC~\cite{Pilati10,Conduit09},
demonstrating the quantitative benefits of using a high fidelity
pseudopotential. The presence of the first order transition is consistent
with theory~\cite{Duine05,Conduit09,Maslov09,Pedder13} and with
the ferromagnetic transition seen in experiments on heavy fermion materials~\cite{Belitz12}.

\emph{P-wave superconductivity} can be understood by considering two up-spin electrons in a
fermionic gas with repulsive interactions, each surrounded by a fluctuating
magnetic polarization cloud. As the electrons coalesce the magnetic
fluctuations (that drove the first order ferromagnetic transition) reinforce to
create an effective attractive interaction, inducing p-wave superconducting
order~\cite{Hattori12,Julian12}.  The p-wave superconducting state has been
observed in experiments on ferromagnetic
superconductors~\cite{Saxena00,Huxley01,Watanabe02,Aoki01,Huy07,Gasparini:2010,Hattori12},
and has been modeled by a contact interaction in perturbation
theory~\cite{Fay:1980kx,Mathur1998,Roussev:2001ys,Kirkpatrick:2001vn,Wang:2001kl,Conduit13},
but has never been observed in numerics. Equipped with a pseudopotential that
reproduces the contact interaction with high fidelity and whose broad profile
leads to improved efficiency, we search for p-wave superconducting order.

The p-wave superconducting order is defined by the order parameter
$\Delta_\vec{k}=\sum_{\vec{k}'} V_{\vec{k}\vec{k}'}\langle
c_{\vec{k}\uparrow}c_{-\vec{k}\uparrow}\rangle$. This must be recast into an
operator in the position representation and projected onto the p-wave
channel. Effecting this transformation results in the projection of the
off-diagonal long-range order in the two-body reduced density matrix onto
the p-wave channel~\cite{Yang62,Wagner13,Morris10}
\begin{align}
    \langle|\Delta_\mathrm{p}|^2\rangle \!=\!-\frac{(4\pi\kf a)^2}{81\Omega^2}
    \!\lim_{\substack{\Pos\\\to\!\infty}}\!\iint\!\!
    \pos\!\cdot\!\pos'\langle \cre_\frac{\pos}{2}
    \cre_{-\!\frac{\pos}{2}} \ann_{\Pos\!+\!\frac{\pos'}{2}}
    \ann_{\Pos\!-\!\frac{\pos'}{2}}\rangle \diffd\pos \diffd\pos'\nonumber\!,
\end{align}
where $\Omega$ is the simulation cell volume.  The expectation value is zero
for the Slater determinant trial wavefunction $D_{\uparrow}D_{\downarrow}$
with no electron-electron correlations. However, if we insert the full trial
wavefunction $\psi=\e{J}D_{\uparrow}D_{\downarrow}$ into the expectation value
and expand in the limit of small electron separation,
we find that
$\langle|\Delta_\mathrm{p}|^2\rangle\approx2^{10}3^{-15}5^{-2}7^{-1}(k_{\text{F}}c)^{8}u_{\uparrow\uparrow}(0)$,
connecting the superconducting correlations to the up-spin correlation term
in the Jastrow factor. This verifies that the trial wave function has the
variational freedom to exhibit a superconducting instability.

In \figref{fig:ferro} we show the emergence of the p-wave superconducting
order parameter with increasing interaction strength. The p-wave
superconductor may be enhanced in the partially polarized
phase~\cite{Conduit13}, but is destroyed in the fully polarized state as there
can be no magnetic fluctuations. The
delicacy of the superconducting order requires a high-fidelity pseudopotential. The emergence
of the p-wave superconducting order provides the first verification of the
magnetic fluctuations theory valid at high interaction strengths,
confirming the NMR measurements on UCoGe~\cite{Hattori12}.

\section{Discussion}

We have developed a high fidelity pseudopotential for the contact
interaction. The pseudopotential is ultratranserable, delivering accurate
scattering properties over all wavevectors $0\le k\le2\kf$ in a Fermi gas and
its smoothness accelerates computation.
This pseudopotential allowed us to characterize the first order
itinerant ferromagnetic transition and present the first computational
evidence for a p-wave superconducting state.

The performance and portability of the pseudopotential makes it widely
applicable
across first principles methods including VMC, DMC,
coupled cluster theory, and configuration interaction. The formalism
developed can also be applied more widely in scattering problems in
condensed matter to develop pseudopotentials, including the repulsive Coulomb
interaction and dipolar interactions.

\acknowledgments{The authors thank Stefan Baur, Andrew Green, Jesper Levinsen, Gunnar
  M\"oller, Michael Rutter, and Lukas Wagner for useful discussions, and acknowledge the
  financial support of the EPSRC and Gonville \& Caius College.}

\end{document}